\documentclass[fleqn,usenatbib]{mnras} 

\usepackage{newtxtext,newtxmath}

\usepackage[T1]{fontenc}
\usepackage{ae,aecompl}

\usepackage{graphicx}	
\usepackage{amsmath}	
\usepackage{amssymb}	

\def\subsun{\mbox{$_{\normalsize\odot}$}}

\def\farcs{\hbox{$.\!\!^{\prime\prime}$}}

\def\micron{\hbox{$\mu$m}}

\def\VY{\hbox{VY~CMa}}

\def\pipe{\hbox{\sc precision}}

\title[]{{\sc precision}: A fast python pipeline for high-contrast imaging -- application to SPHERE observations of the red supergiant VX Sagitariae\thanks{Based on observations collected at the European Organisation for Astronomical Research in the Southern Hemisphere under ESO programmes 095.D-0656.}} 

\author[P. Scicluna et al.]{P. Scicluna,$^{1\thanks{E-mail: peterscicluna@asiaa.sinica.edu.tw (PS)}}$
F. Kemper,$^{2,1}$
R. Siebenmorgen,$^2$
R. Wesson,$^3$
J.A.D.L. Blommaert,$^4$
\newauthor
S. Wolf\,$^5$
\\
$^{1}$Academia Sinica, Institute of Astronomy and Astrophysics, 11F Astronomy-Mathematics Building, NTU/AS campus, No. 1, Section 4, \\Roosevelt Rd., Taipei 10617, Taiwan\\ 
$^2$European Southern Observatory, Karl-Schwarzschild-Str. 2, 85748, Garching b. M\"unchen, Germany \\
$^3$Dept. of Physics \& Astronomy, University College London, Gower Street, London WC1E 6BT, UK \\
$^4$Astronomy and Astrophysics Research Group, Dep. of Physics and Astrophysics, V.U. Brussel, Pleinlaan 2, 1050, Brussels, Belgium \\
$^5$Institute of Theoretical Physics and Astrophysics, Universit\"at zu Kiel, Leibnizstr. 15, 24118, Kiel, Germany
}

\date{Accepted XXX. Received YYY; in original form ZZZ}

\pubyear{2016}

\begin{document}
\label{firstpage}
\pagerange{\pageref{firstpage}--\pageref{lastpage}}
\maketitle\relax

\begin{abstract}
The search for extrasolar planets has driven rapid advances in instrumentation, resulting in cameras such as SPHERE at the VLT, GPI at Gemini South and SCExAO at Subaru, capable of achieving very high contrast ($\sim10^{6}$) around bright stars with small inner working angles ($\sim 0\farcs{1}$).
The optimal exploitation of data from these instruments depends on the availability of easy-to-use software to process and analyse their data products.  
We present a pure-python pipeline, \pipe, which provides fast, memory-efficient reduction of data from the SPHERE/IRDIS near-infrared imager, and can be readily extended to other instruments.
We apply \pipe\ to observations of the extreme red supergiant VX~Sgr, the inner outflow of which is revealed to host complex, asymmetric structure in the near-IR.
In addition, optical polarimetric imaging reveals clear extended polarised emission on $\sim0.5^{\prime\prime}$ scales { which varies significantly with azimuth, confirming the asymmetry.} {While not conclusive, this could suggest} that the ejecta are confined to a disc or torus, which we are viewing nearly face on, { although other non-spherical or clumpy configurations remain possible}.
VX~Sgr has no known companions, making such a geometry difficult to explain, as there is no obvious source of angular momentum in the system.
\end{abstract}




\section{Introduction}
Since the first detections of planets outside our solar system, the astronomical community has been pushing for ever greater advances in observing methodology to enable the direct imaging and characterisation of these exoplanets.
One particular avenue is the pursuit of ever-higher contrast at ever-smaller angular distances from bright, nearby stars.
This has driven the development of advanced processing algorithms to correct for light-path aberrations or remove the PSF of the central star, but also technological progress.
The combination of advanced coronagraphs, high-order adaptive optics (sometimes referred to as eXtreme Adaptive Optics, hereafter XAO), large-aperture telescopes, high dynamic range, high speed detectors and state-of-the-art differential-imaging techniques has recently given rise to a new generation of planet-hunting instruments, such as SPHERE \citep{2008SPIE.7014E..18B}, GPI \citep{2006SPIE.6272E..0LM} and SCExAO \citep{2009SPIE.7440E..0OM}.
By effectively suppressing the PSF of the star and maximising the sensitivity to additional point sources, these instruments have made numerous discoveries in the few short years they have been in operation \citep[e.g.][and papers citing it]{2016A&A...587A..58B}.

Beyond the primary science cases of exoplanet detection and characterisation, the capabilities of XAO instruments are in high demand.
The combination of high angular resolution, contrast, and sensitivity with { a well-calibrated PSF} has generated advances in fields as diverse as star and planet formation \citep[e.g.][]{Pinilla2015}, debris discs \citep[e.g.][]{2016A&A...596L...4W}, substellar objects \citep[e.g.][]{2016A&A...587A..55V}, binaries and star clusters \citep[e.g.][]{2016A&A...588L...7K}, AGN tori \citep[e.g.][]{2015A&A...581L...8G}, small solar-system bodies \citep[e.g.][]{2015A&A...581L...3V,2016ApJ...820L..35Y}, evolved stars \citep[e.g.][]{Kervella2015,2015A&A...584L..10S,2016A&A...585A..28K,Ohnaka2016} and stellar remnants \citep[e.g. white dwarfs,][]{Xu2015}.
Unlike exoplanet searches, many of these science cases are concerned with extended emission, which they wish to image at high angular resolution close to bright central sources.
In the case of evolved stars, for example, XAO instruments are now able to resolve the dust-formation zone, where high-contrast observations of dust-scattered light can reveal details of the mass-loss and dust-formation processes. 
The differential-imaging methods commonly employed for exoplanet imaging, however, will partially or completely self-subtract extended continuum emission \citep[e.g.][]{2014ApJ...780...25E}. 

Given the specialised nature of the instruments, the data they produce and the processing required to optimally exploit them, data reduction and analysis pipelines are now a key part of the scientific process for these instruments.
This is part of a growing trend of software becoming as important to the scientific process as the data it is used to analyse. 
The rise of highly-specialised data-reduction software presents a problem, however, as it can be prohibitively difficult to optimise if one is not already an expert, and many remain outside the public domain, preventing public scrutiny of the code.
Furthermore, the advanced processing techniques required by high-contrast imaging necessitate proscriptive amounts of computing power and manual intervention.
As a result, automated reduction and batch processing are problematic.

To alleviate some of these issues, we have developed a highly-automated data reduction pipeline optimised for the analysis of bright targets and extended emission.
This pipeline, \pipe, is presented in Sect.~\ref{sec:prec}, and is designed to provide a fast, simple and memory-efficient way to process high-contrast imaging data and produce science-ready products while still leaving routes open to perform more advanced, time-consuming reductions at a later date.
Furthermore, it is designed to be easy to use, allowing non-experts to explore their data without being overwhelmed by the available options, and is fully open source.
We then present observations of the galactic extreme Red Supergiant (RSG) VX~Sgr as a test case to demonstrate the use and performance of \pipe.
The observations are presented in Sect.~\ref{sec:obs}, and analysed in Sect.~\ref{sec:res}.

\section{IRDIS data reduction with {\sc precision}}\label{sec:prec}
We present the new infrared imaging pipeline {\sc precision}, which is a publicly available, open source python package licensed under the GPLv3 and available to the community at \url{https://github.com/pscicluna/precision.git}. 
{\sc precision} is written in pure python, using the Numpy \citep{numpy}, Scipy \citep{scipy}, astropy \citep{astropy_paper,astropy_ascl} and photutils \citep{photutils} packages to provide key functionality.
While the current version only implements features for the reduction of {\sc SPHERE/IRDIS} data, it is designed to be flexible and extensible, and it is expected that future versions will implement pipeline objects for other infrared imagers such as NACO and VISIR.

The design philosophy is intended to provide a high degree of automation and to reuse as much code as possible within the package, while still making it possible to manually control or overload\footnote{replace methods bound to the pipeline objects with functions written by the user if the desired functionality is not included in \pipe} the data reduction process if so desired.
To facilitate this, the package includes modules to identify and parse ESO association trees \citep{2013A&A...559A..96F} to resolve the relationships between the science and calibration data and apply the correct calibrations to each frame.
It also aims to provide a fast\footnote{A typical observation with $\sim$ 500 frames can be reduced from start to finish in a few minutes.}, memory-efficient\footnote{For a typical observation, the peak RAM usage is in the region of a few hundred MB.} reduction while still producing science-ready products.
This makes it an excellent tool for quick-look analysis, batch processing or reduction on lower-spec computers e.g. laptops.
{ However, as the emphasis is on speed, automation and portability, it is not expected to provide optimal products for all use cases. }

\begin{figure*}
    \centering
    \includegraphics[width=1.3\textwidth,angle=90]{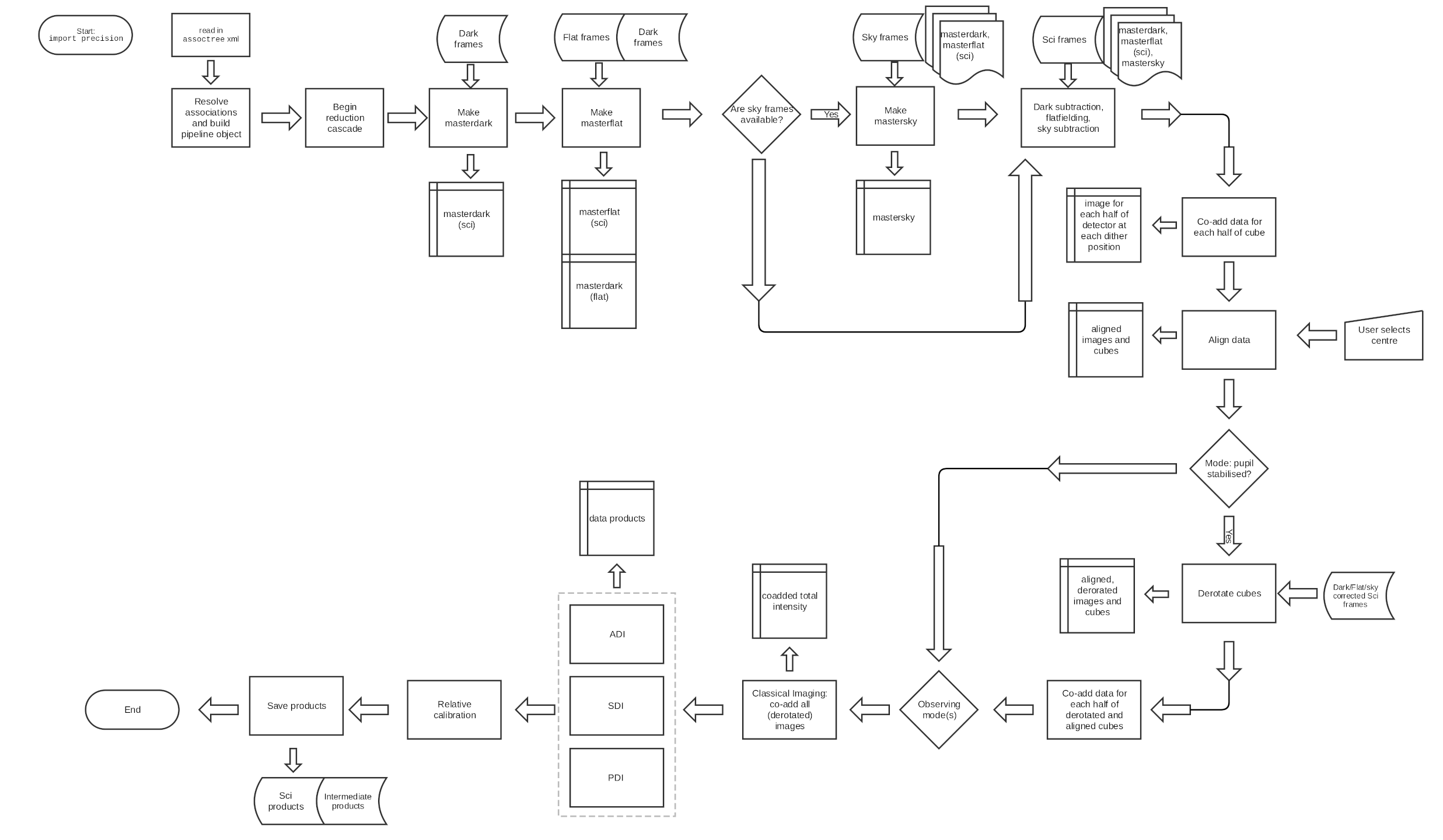}
    \caption{Flowchart depicting the reduction cascade employed by {\sc precision}.}
    \label{fig:flowchart}
\end{figure*}

{\sc precision} implements a standard workflow for {\sc IRDIS} observations, consisting of (i) dark subtraction (ii) flat fielding (iii) sky subtraction\footnote{\label{ifposs}If appropriate calibration frames are available} (iv) co-addition (v) de-rotation \& co-addition of de-rotated frames\footnote{Only applicable to pupil-stabilised observations} and (vi) relative flux calibration\textsuperscript{\ref{ifposs}}. 
In addition, if the data being reduced supports it, {\sc precision} will also perform angular differential imaging (ADI), spectral differential imaging (SDI), polarimetric (differential) imaging (PDI) and reference-PSF subtraction (reference differential imaging or RDI). 
The approaches taken in these steps are described in detail below, with the overall flow shown in Fig.~\ref{fig:flowchart}. 

The routine begins by constructing a data-reduction object (\emph{pipeline}) for each group of files to be reduced together. 
This group should consist of a series of objects containing references to each data file and the full cascade of calibration files required for use in the pipeline.
For ease, these can be automatically constructed directly from ESO association tree xml files if available.
Once the pipeline has built the reduction cascade, it can be instructed to automatically proceed through all of the following steps that the observing mode supports.
Alternatively, the user can activate specific steps of the reduction process interactively or by scripting the reduction, to allow other parts to be overridden.

\subsection{Mandatory reduction steps}
\subsubsection{Dark subtraction}\label{sec:dark}
For any group of dark frames $D_{i = 0,\dots,n}$, a master dark $\mathbf{D}$ is constructed by a simple procedure by
first dividing by the exposure time to convert to counts~s$^{-1}$, and then median-combining all frames, i.e.
\begin{equation}
    \mathbf{D} = \Tilde{D},
\end{equation} where $\Tilde{D}$ indicates the median of $D_{i = 0,\dots,n}$ along the time axis.
The uncertainty on each pixel is computed using classical statistics as $\sqrt{\frac{2}{\pi}} \sigma_{D_i}$ when multiple frames are combined, where $\sigma_{D_i}$ is the standard deviation in each pixel of the group of frames.
If only a single frame is used, the uncertainty is assumed to be proportional to the square root of the absolute pixel values. 
Unless otherwise specified, these two methods are applied to derive uncertainties in all further steps where relevant.
This master dark is then subtracted from the observations with which it is associated.

\subsubsection{Flat fielding and bad-pixel correction}\label{sec:flat}
After dark current has been removed, it is necessary to correct for any non-uniformity of the response of the detector.
For each flatfield observation, the corresponding master dark is built by the process outlined above, and subtracted from each flat frame $F_{i = 0,\dots,n}$.
Each dark-subtracted flat frame is then divided by its median value, and all flat frames are combined by taking the median value of each pixel to produce the master flat $\mathbf{F}$, i.e. 
\begin{align}
    F_{i = 0,\dots,n}^{\prime} &= F_{i = 0,\dots,n} - \mathbf{D} \\
    F_{i = 0,\dots,n}^{\prime\prime}   &=  \frac{F_{i = 0,\dots,n}^{\prime}}{\mathrm{median}\left(F_{i = 0,\dots,n}^{\prime}\right)} \\
    \mathbf{F} &= \Tilde{F^{\prime\prime}}
\end{align}
This master flat is used to correct the frames with which it is associated by dividing their pixel values by those in the master flat.

Bad pixels are then identified from the master flat by sigma clipping.
For the IRDIS 1K$\times$2K detector, one expects $\sim$1 pixel more than 5$\sigma$ away from the mean under the assumption that the pixel values are normally distributed.
\pipe~therefore assumes that any pixel that deviates by 5$\sigma$ or more is a bad pixel, and includes it in the bad-pixel map, although the threshold can be modified.
In subsequent reduction steps, these pixels are given the value ``not a number'' (NaN) in other frames, and the astropy routine \emph{convolution.interpolate\_replace\_nans} is used to correct these pixels.

\subsubsection{Frame alignment and combination}\label{sec:align}
In order to facilitate differential-imaging techinques, IRDIS images the same field onto each 1k$\times$1k half of the detector.
It is therefore necessary, in addition to correcting for dithering, to separate the two halves of the detector and combine both images.

Following dark subtraction, flat fielding, bad-pixel correction (and optionally sky subtraction, see below), the data cubes are shifted to correct for dithering offsets.
Then, the cubes are separated into their two halves, $I_{\rm L}$ and $I_{\rm R}$, and aligned so that the optical axis (whose location is determined as in Sect.~\ref{sec:cent}) is at the centre of the image.
The two cubes are then concatenated along the time axis and combined by taking the median of the pixel values along this axis.
This is repeated for all available dither positions, and the resulting $N_{\rm dither}$ images are combined by taking the mean of the pixel values at all dither positions.

\subsection{Optional steps}\label{sec:opt}
\subsubsection{Sky subtraction}\label{sec:sky}
If sky background observations have been taken along with the science observation, it is necessary to subtract this background from the science data.
In \pipe~this is implemented as follows
\begin{enumerate}
    \item for each frame of the sky observation, perform
    \begin{enumerate}
        \item dark subtraction,
        \item flat fielding,
        \item interpolation of bad pixels; then 
    \end{enumerate}
    \item median-combine all frames of the sky observation
\end{enumerate}
to produce one master sky frame $\mathbf{S}$ i.e.
\begin{align}
    S_{i = 0,\dots,n}^{\prime} &= \frac{S_{i = 0,\dots,n} - \mathbf{D}}{\mathbf{F}}\\ 
    \mathbf{S}  &= \Tilde{S^{\prime}}
\end{align} which can then be subtracted from each science frame.
This subtraction is performed after dark subtraction and flat fielding, but before frame alignment.

\subsubsection{Centring and derotation}\label{sec:cent}
SPHERE supports both field-stabilised observations, where the field-of-view is fixed in the detector frame throughout the observation, and pupil-stablised, where the orientation of the pupil is fixed with respect to the detector and hence the field-of-view rotates.
In this case it is therefore necessary to derotate the images to a common frame of reference, which in turn requires the optical axis to be determined precisely.
In the case of SPHERE, the optical axis is aligned with the AO guide star, which in the case of coronographic observations is obscured by the coronagraph. 
The centre of the coronagraph is indicated by a bright spot created by diffraction, and if the coronagraph is correctly centred this is also the position of the star.
\pipe\,therefore relies on determining the position of this central spot; while less precise than other centring algorithms e.g. the use of a waffle pattern on the deformable mirror to project multiple satellite images of the source at well-defined offsets, the ease and speed of this method makes it more amenable to the objectives of \pipe.
Future development will also include the determination of the optical axis from the satellite spots for cases where centring is of particular importance.

To facilitate this, we use the co-added, sky-subtracted images of each half of the detector at each dither position.
For each of these images, we use the photutils {\sc DAOStarFinder.find\_stars} implementation to locate all point-like sources within a user-adjustable window of the detector-position of the coronograph at the last determination of its position (as recorded in the FITS header).
The user is then presented with a table of source properties and an image of the search window, and required to input the ID of the source they believe to be the centre of rotation.
The centroid of this source is then used to align the image so that this source is centred in the middle of the array, and the images are derotated about their centre.
At present, this is the only reduction step that requires human interaction and all other steps can be fully automated; we are looking for ways of automating this step to allow for fully scriptable data reduction.

Centring is probably the largest source of systematic uncertainty in \pipe. 
Failure to correctly determine the centre of rotation will result in smearing of both real extended emission, reducing its significance and making it appear more extended, and of speckles.
Smearing of speckles may introduce artefacts that simulate extended emission, similar to failing to derotate the data or to the smearing of the PSF in field-stabilised mode.
In addition, it will have a significant adverse effect on the contrast achievable when using Angular Differential Imaging (see Sect.~\ref{sec:cadi}).
As a result, the magnitude of this effect will depend strongly on the amount of field rotation during the observation.
However, as VX~Sgr transits at zenith from Paranal, the impact on the observations presented in Sect.~\ref{sec:obs} is negligible, since the total field rotation is small.
{ The next item of concern is that the constituent cubes (one for each half of the detector at each dither position) were not aligned correctly prior to collapse. However, this would again be quite obvious as the coronagraph would appear non-circular, and the central spot significantly smeared and different from each individual cube. To mimic extended emission it would have to be offset by several PSF-widths, which would produce a set of obvious rings from the coronagraph, making it easy to identify, while the typical dither offsets are one or two pixels.}
These issues will be explored in more detail in the documentation and tutorials for {\sc precision}.

\subsubsection{Relative flux calibration}\label{sec:rfc}
In coronographic imaging, the central source is hidden or strongly attenuated. 
To determine the contrast scale it is therefore necessary to take additional images where the central source is not obscured.
\pipe~takes these images and reduces them by dark subtracting, flat fielding and sky subtracting them. 
It then combines all the frames if more than one is available, locates the brightest point source near the centre of the field and scales the coronographic images by dividing by the peak flux of this source.
If the flux observations were taken using a neutral-density filter, it rescales the peak flux by dividing by the throughput of the neutral-density filter at that wavelength, computed by convolving the filter-response curves with the throughput curve of the relevant neutral-density filter.
A table of throughputs is distributed with the code.

\subsubsection{PSF subtraction}\label{sec:psfsub}
The major limitation to the contrast that can be achieved with XAO systems is so--called ``speckle noise''.
This arises because of quasi-static aberrations in the optical path which result in faint copies of the PSF core (``speckles'') being distributed across the pupil plane.
In order to remove these, it is necessary to have a well-calibrated reference PSF to subtract from the science data, the simplest case of which is Reference Differential Imaging (RDI), in which a second source close on the sky, which is expected to be point-like, is observed immediately after the science target.
\pipe~includes the option to associate two sets of observations together, with one serving as the reference PSF which will be subtracted from the other.
The two datasets are reduced independently, then the final images are aligned and the reference data are scaled so that the central spots have the same peak counts.
The reference data can then be subtracted from the science data to suppress the speckle noise and improve contrast within the AO control radius.

\subsubsection{Classical ADI}\label{sec:cadi}
In pupil-stablised observations, the field-of-view rotates on the detector, while the pupil, and hence the PSF, remains in a fixed orientation on the detector.
This opens a number of options for advanced PSF-subtraction in which the reference PSF is built from the science observations themselves, one particularly successful example of which has been Angular Differential Imaging \citep[ADI,][]{2006ApJ...641..556M}.
\pipe~currently implements the simpler ``classical'' ADI (cADI) algorithm \citep{2006ApJ...641..556M}, which is faster and easier to implement than more advanced methods \citep[e.g. LOCI and MLOCI,][]{2007ApJ...660..770L,2015A&A...581A..24W} at the expense of lower speckle rejection.
This choice is a result of the focus of \pipe~on speed.

To perform the cADI reduction, the dark- and flat-field corrected frames are aligned before derotation, and co-added to produce a single PSF image.
For each frame, this image is subtracted to remove the PSF; each frame is then derotated to a common frame and combined to produce a single image from which the static structure of the PSF has been removed, but real structures (which rotate with the FOV in pupil-stabilised observations) are retained provided the field rotation is large enough.
This method also self-subtracts extended emission \citep{2012A&A...545A.111M,2014ApJ...780...25E}, improving the contrast of companion candidates or small-scale structures in the stellar winds.

\subsubsection{SDI}
Spectral differential imaging exploits the existence of differences between the spectral features of different classes of objects, primarily between stars and substellar objects.
If the spectral feature is compared to sufficiently-close (in wavelength) continuum, the PSF barely changes between the two images, allowing the continuum to be used to efficiently subtract the PSF from the emission in the spectral feature \citep{HaydenSmith1987}.
Since IRDIS exploits special filters to project the on-feature and off-feature beams onto the two halves of the detector, it is sufficient to subtract the reduced, median-combined images of each half of the detector instead of co-adding them, i.e.\begin{equation} I_{\rm SDI} = I_{\rm L} - I_{\rm R},\end{equation}
where the subscripts $L$ and $R$ correspond to the left and right halves of the detector, respectively.
\pipe~will also attempt to combine SDI and ADI if the data supports it, i.e. if an SDI filter is combined with pupil-stabilised imaging.

\subsubsection{PDI}
{\sc precision} also includes routines to handle polarimetric imaging. 
Polarimetry is particularly interesting for observing extended, dusty structures, as scattering of the (initially unpolarised) stellar light by the dust grains introduces strong polarisation.
IRDIS facilitates polarimetry by employing a rotatable half wave plate (HWP) and a Wollaston prism as a polarising beam splitter.
This results in the Ordinary (O-ray) and Extraordinary (E-ray) rays each being imaged onto one half of the detector. 
Rotating the HWP then changes the polarisation state that is sampled.
To image both the Q and U components of the Stokes' vector, HWP angles $(\phi)$ of 0$^\circ$ and 22.5$^\circ$ are required, and further rotations of 45$^\circ$ and 67.5$^\circ$ allow for correction of systematic effects which cancel out in the two orthogonal polarisation frames of reference.
In \pipe~this is implemented with both the double-ratio \citep{1996aspo.book.....T,2006A&A...452..657S} and double-difference methods \citep{2009ApJ...701..804H}, similar to \citet{2011ApJ...738...23Q}. These methods are described in more detail below.

Before polarimetric processing, all frames are dark-subtracted, flat fielded, bad-pixel corrected, aligned and co-added so that there are two images for each HWP position angle, one for each half of the detector. 
These frames can then be combined to produce $(I, Q, U)$ images, or $(I, p, \theta)$. 
The total intensity $I$ is trivially determined by first summing the two halves of the detector at each $\phi$ and then taking the average over all $\phi$.

$Q$ and $U$ are determined by the double-difference method.
This relies on the fact that each $\phi$ corresponds to one Stokes' parameter so that $\phi = \left(0, 22.5,45, 67.5\right)$ traces $\left(+Q, +U, -Q, -U \right)$ respectively.
For each $\phi$ one therefore calculates e.g. $$+Q = O_0 - E_0 = I_{L,0} - I_{R,0} $$ where O corresponds to the intensity of the ordinary-ray and E the extraordinary-ray when $\phi=0$.
This is applied to all the pairs of $\phi$ and Stokes' parameter given above, and the pairs of $+/-~Q/U$ are then combined to give 
\begin{align}
Q &= \left(+Q - \left( -Q \right)\right) / 2 \\
U &= \left(+U - \left( -U \right)\right) / 2
\end{align}
and from these $Q$ and $U$ images, $p$ and $\theta$ are calculated in the normal way i.e.
\begin{align}
    p &= \frac{\sqrt{Q^2 + U^2}}{I} \\
    2\theta &= \tan^{-1}\left(\frac{U}{Q}\right). 
\end{align}

The double-ratio method calculates the fractional Stokes' parameters $p_Q$ and $p_U$.
It begins by calculating
\begin{align}
    R_Q &= \sqrt{\frac{O_0 / E_0}{O_{45} / E_{45}}}\\
    R_U &= \sqrt{\frac{O_{22.5} / E_{22.5}}{O_{67.5} / E_{467.}}}
\end{align}
to cancel out the influence of the atmosphere and other time-dependent factors.
From this, the fractional Stokes' parameters a
\begin{equation}
    p_Q = \frac{R_Q - 1}{R_Q + 1}
\end{equation}
and similarly for $p_U$. 
Finally, similar to the double-difference method, the fractional polarisation is determined from
\begin{equation}
    p = \sqrt{p_Q^2 + p_U^2}.
\end{equation}

\subsection{Benchmarking}{
To briefly demonstrate the performance of \pipe\ and compare it with other data-reduction pipelines, we reproduce the {\it noADI} reduction from \citet{Langlois2018}, who performed SDI on the transition disc RY Lupi.
In their Figure 1, \citet{Langlois2018} showed the result of combining both spectral channels with derotation, where the centre of rotation was derived by using four satellite spots in a cross pattern.
This allows us to demonstrate the difference in performance of the centring when the field rotation is large, and hence the accuracy of the centring is most critical.

\begin{figure}
    \centering
    \includegraphics[width=0.45\textwidth]{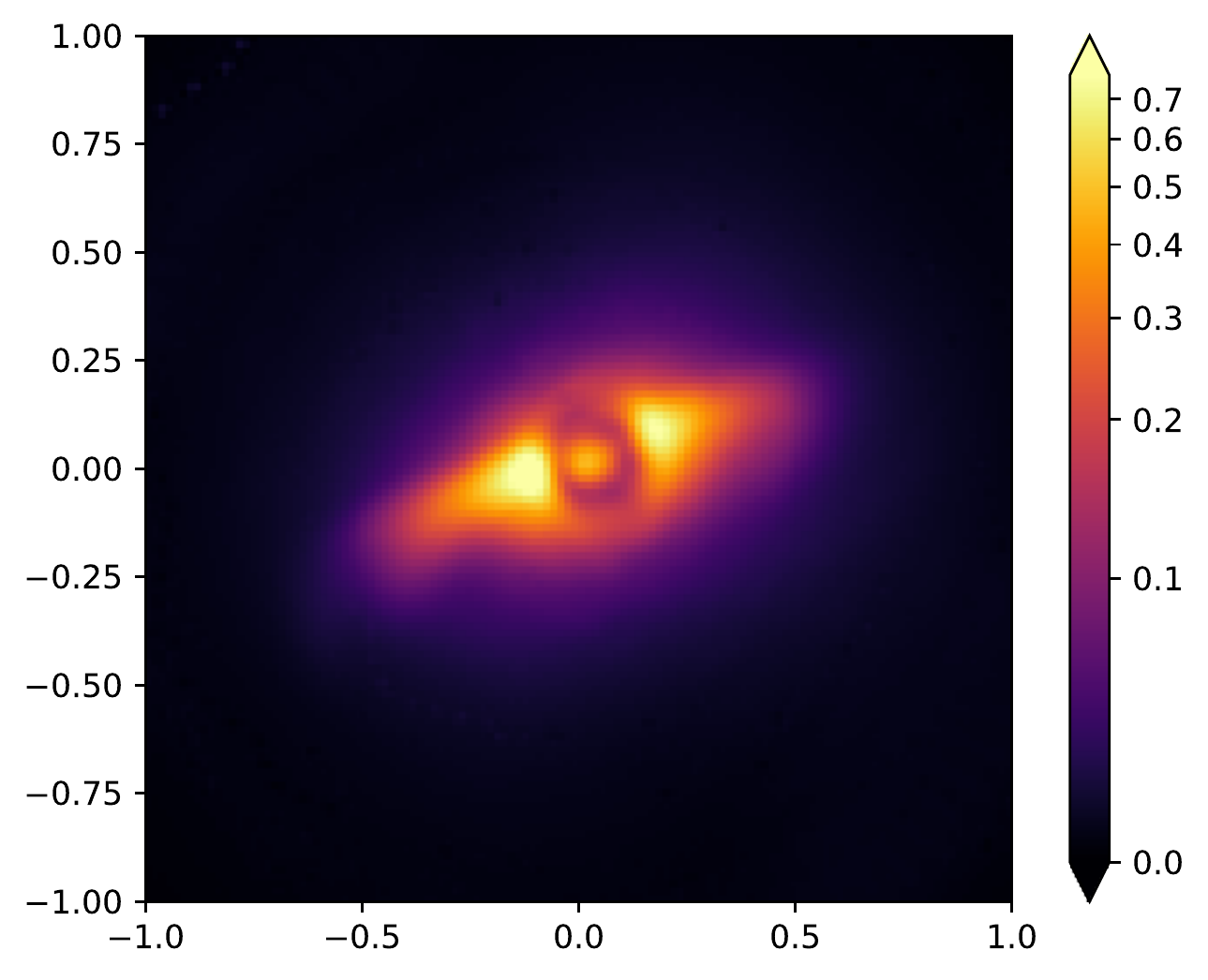}
    \caption{\pipe\ reduction of the RY Lupi data presented in \citet{Langlois2018}, with N up and E to the left shown on an asinh scale normalised to the brightest pixel in the image. The birfurcation of the eastern part of the disc is less clearly visible here than in \citet{Langlois2018}, but \pipe\ otherwise reproduces the features seen in the literature.}
    \label{fig:bench}
\end{figure}

Our reduction of the same dataset included the usual steps of dark subtraction, flat fielding, sky subtraction, and centring and derotation. 
As we are primarily concerned with whether it reproduces the published morphology as a test of the centring, we do not worry about flux calibration or PSF subtraction.
The entire reduction takes roughly 10 minutes on a commercial laptop computer, providing easy reduction without needing to resort to data-centre processing.
The resulting image can be seen in Fig.~\ref{fig:bench}.
Comparison with the top right-hand panel of Figure 1 in \citet{Langlois2018} shows that \pipe\ generally reproduces the same features.
There is some blurring of the fine structures of the disc, a result of the reduced centring accuracy, but this is relatively small.
For many use cases, this will be acceptable.
}
\section{Demonstrating \pipe\ on the extreme red supergiant VX~Sgr}

\subsection{Background}
After massive stars (M~$>$~8~M\subsun) leave the main sequence, they experience several episodes of enhanced mass loss before exploding as supernovae. 
The scale and nature of this mass loss is a key factor in determining the evolution of the stars through phases including Red Supergiant (RSG), Yellow Hypergiant (YHG), Luminous Blue Variable (LBV), and Wolf-Rayet (WR) \citep{2013EAS....60...43G,2014ARA&A..52..487S,2015A&A...575A..60M,2015EAS....71...41G}. 
Moreover, if a star explodes within a dense mass-loss envelope, the effect on the resulting supernova is dramatic; the interaction of the explosion with the ejecta releases huge quantities of energy and triggers rapid dust formation \citep{2010ASPC..425..279V}.
A proper understanding of mass loss is therefore crucial for determining the post-main sequence evolution of massive stars and for linking classes of supernova progenitors to classes of supernovae \citep{2012A&A...538L...8G,2013A&A...550L...7G,2013EAS....60...31E,2013EAS....60...43G,2013EAS....60..307V}.

For stars with initial masses $\lesssim30$\,M$_\odot$, a large fraction of the mass loss occurs during the RSG phase, but this process remains poorly understood.
Seemingly typical RSGs (e.g. $\alpha$~Ori) have mass-loss rates orders of magnitude too low compared to evolutionary models \citep{2013EAS....60..307V}, while `extreme' RSGs such as VY~CMa and NML~Cyg are losing mass far too quickly, with evidence of short-term {\it eruptive} increases in mass-loss as well. 
Furthermore, recent high-contrast observations of $\alpha$~Ori have shown that lower mass-loss rate RSGs also have highly inhomogeneous outflows, and that the inhomogeneity begins within a few stellar radii \citep{2009A&A...504..115K,2011A&A...531A.117K,2016A&A...585A..28K}.

The mechanisms driving RSG mass-loss remain a matter of debate \citep{2005A&A...438..273V,2009ApJ...701.1464H}, in particular the origins of mass-loss asymmetries, variability and eruptions.
Various mechanisms, which may not necessarily be independent, have been invoked, including non-radial pulsations, convection, binarity, and magnetic activity \citep[e.g.][]{2001AJ....121.1111S,2007AJ....133.2716H}.

Understanding the mass-loss process and the mechanisms driving it are also key to understanding the connections between different classes of evolved massive stars and the supernovae they become \citep{2010ASPC..425..279V,2014ARA&A..52..487S}.
Enhanced mass loss will alter the internal balance of the star, producing blueward evolution in the Herzsprung-Russell (HR) diagram, potentially multiple times \citep[so-called "blue-loops", e.g.][]{1999A&A...342..131S,2012A&A...537A.146E}, influencing the relative size of the red, yellow and blue supergiant populations. 
It may also produce a dense envelope of material with which the supernova will interact, producing exotic lightcurves, or strip enough mass to influence the type of compact remnant left behind.

VX~Sgr is a nearby (1.5\,kpc, \citealt{2007ChJAA...7..531C}) dusty RSG with a mass-loss rate $\sim 6 \times 10^{-5}$\,M\subsun\,yr$^{-1}$ and luminosity~$\sim~10^6$\,L\subsun\,\citep{2010A&A...523A..18D}.
This luminosity places it near the empirical upper limit for luminosity \citep{1979ApJ...232..409H}; stars in this region of the HR diagram are typically associated with eruptive mass loss and variability.
Near-IR interferometry by \citet[][]{2010A&A...511A..51C} show that the photospheric radius is in the range 2.5--4.5 mas depending on wavelength, but that the star has several bright spots that probably form in the molecular layers. 
It is a powerful maser source, the polarisation of which has been used to trace large scale dipole magnetic fields in the envelope, whose strengths imply surface fields of a few Gauss \citep{2011ApJ...728..149V}.
The morphology of the 7~mm SiO maser emission clearly traces a ring on ~1-10\,mas scales \citep{2012ApJ...754...47S}; the coherent velocity structure of this ring is suggestive of an equatorial outflow within 3~$R_{\star}$ seen pole-on.
Water maser emission is highly inhomogeneous and concentrated to the NW \citep{2018arXiv180704455Y}, with significant ellipticity. 
It has previously been suggested that this is consistent with a spheroidal shell inclined by $\sim 45^{\circ}$ from the line of sight \citep{1996PhDT........82M}.
\citet{2005A&A...434.1029V} found evidence for a large-scale, ordered, dipole magnetic field with an inclination of 40\,$\pm$\,5\,$^\circ$, which may be related to the asymmetry.
Recent results from \citet{2018AJ....155..212G} present contradictory pictures of the recent evolution of VX~Sgr. 
They fit the SED using spherical-outflow models computed with {\sc dusty}, and the best-fitting models imply that the mass-loss rate was higher in the past. 
However, comparing the predicted mid-infrared radial profiles of these models to SOFIA/FORCAST imaging (their Figure 3) strongly contradicts this scenario, with the lack of resolved emission in the profiles at 19.7 and 25.3\,\micron\,in fact implying that the present-day mass-loss rate is higher.
This contradiction, combined with the excess of emission in the 5 -- 10\,\micron\,range compared to all the \citet{2018AJ....155..212G} models implies that, similar to the maser emission, the dust geometry is more complex than these models are able to describe, making it an ideal target for high-contrast imaging.

The high mass-loss rate and luminosity places VX~Sgr in the small group of "extreme" RSGs, along with VY~CMa and NML~Cyg. 
It has been suggested that these sources are near the end of the RSG phase, and that their high mass-loss rates may presage blue-ward evolution, making them the progenitors of YHGs or WRs \citep{2009AJ....137.3558S}.
Alternatively, if they have initial masses below $\sim$~20\,M$_\odot$, we may be observing a final burst of pre-supernova mass loss \citep{2009AJ....137.3558S}.
As a result, sources in this phase are key to constraining the late phases of the evolution of massive stars, and understanding the process by which they lose mass is critical. 

\subsection{Observations}\label{sec:obs}

\begin{table}
\caption{Observing conditions}
    \centering
    \begin{tabular}{lrr}
    \hline
         & ZIMPOL & IRDIS \\\hline\hline
        Seeing in V band (\arcsec) & 1.2 & 1.2 \\
        Wind speed (m\,s$^{-1}$) & 5 & 10\\
        Transparency & Clear & Clear\\
        DIT (s) & 10 (V band), 40 (Cnt820) & 0.837\\
        NDIT &8 (V-band), 2 (Cnt820) & 100\\
        NDITH & 3 (per polarisation) & 4 \\
        \hline
    \end{tabular}
    \label{tab:conds}
\end{table}

\subsubsection{Optical imaging polarimetry with ZIMPOL}

VX~Sgr was observed on 2015-09-11 using V-band and 820\,nm-continuum(Cnt820)  filters with the Zurich IMaging POLarimeter \citep[ZIMPOL,][]{2008SPIE.7014E..3FT,2014SPIE.9147E..3WR} sub-instrument in active polarisation compensation (p2) mode, which allows for both high-precision ($\sigma_{\rm p_l} \sim 0.3 \%$) polarimetry and long ($\sim 1$\,h) observing sequences to capture the ordinary and extraordinary rays at 4 position angles on sky. 
To avoid saturation and increase contrast, the observations were conducted using a coronagraph to suppress the stellar PSF.
Each polarisation state was observed with a 3-point dither pattern, and 
the science observations were followed by short exposures in which the central star was offset from the coronagraph to allow for flux and contrast measurements; if necessary, a neutral density filter was introduced for these frames to avoid saturation.
The use of polarimetric imaging allows us to suppress the residual PSF halo resulting from imperfect AO correction or the AO control radius as the stars are at most weakly polarised, while scattering by circumstellar dust induces large polarisation of at least several percent.

These observations are identical to those presented in \citet{2015A&A...584L..10S} for VY~CMa, with the exception that the long-wavelength filter was changed from a narrow I-band filter to the narrower Cnt820 filter to provide a more consistent level of exposure in both V and I bands.
Using the same procedure as \citet{2015A&A...584L..10S}, the data were reduced using the ESO instrument pipeline v 0.15 in the {\sc gasgano} environment. 
The data were bias-corrected, flat-fielded and combined to produce Stokes' Q and U images using the double-difference method of polarimetric differential imaging (PDI) \citep[e.g.][]{2001ApJ...553L.189K,2011ApJ...738...23Q}.
These were then combined to give total and linearly polarised intensity and the polarisation angle.

\subsubsection{Near-infrared imaging with IRDIS}

We observed VX~Sgr on 2015-06-11 in the near-IR using the InfraRed Dual-band Imager and Spectrograph \citep[IRDIS,][]{2008SPIE.7014E..3LD} sub-instrument in classical imaging (CI) mode.
Narrow-band continuum $J-$ and $H-$band filters were used to avoid saturation, and a 4-point dither pattern to mitigate the effect of bad pixels.
As with the ZIMPOL observations, a coronagraph was used to suppress the central point source in the science observations.
The full observing sequence in each filter consisted of observations of an empty patch of sky first (\textit{Sky} or S frames), followed by the science observations themselves (\textit{Object} or O), and finally an observation in which the target is offset from the coronagraph to allow the contrast to be determined (\textit{Flux}, F); a neutral density filter was introduced for flux observations to avoid saturation.

Furthermore, the IRDIS observations were followed immediately by short observations of a bright, isolated star 70~Sgr, to remove artefacts from the coronagraphic PSF caused by the finite control radius of the AO system, which were otherwise identical to the observations of the science targets.
To ensure the highest PSF stability, the observations were taken in pupil-stabilised mode, which minimises the change in the optical path for the instrument during the observation.
This also allows the field to rotate on the detector during the observation; provided the integration time is sufficiently short to avoid smearing, the individual frames can then be derotated to the correct orientation on sky and coadded.
This rotation also allows us to exploit more advanced reduction algorithms, such as ADI, 
which enhance the contrast of non-rotationally symmetric structures if the field rotation is large enough (see Sect.~\ref{sec:cadi}).
{ However, as mentioned above VX~Sgr transits at zenith as seen from Paranal, meaning that there is minimal field rotation except during the transit when the source cannot be tracked.
As a result, any uncertainty in the final images injected by imperfect centring should be minimal.
}

These data were automatically reduced using \pipe\,with the associations between science and calibration data determined from the association tree.
The workflow consisted of (in order) dark subtraction (Sect.~\ref{sec:dark}), flat-fielding (\ref{sec:flat}), sky subtraction (\ref{sec:sky}), centring and derotation (\ref{sec:cent}), relative flux calibration (\ref{sec:rfc}) and PSF subtraction (\ref{sec:psfsub}).
Although the data were taken in pupil-stabilised mode to improve PSF stability, ADI is not useful for this dataset as VX~Sgr transits close to zenith at Paranal. 
As a result, the field rotation during the rest of the night is small and self-subtraction is nearly complete.

We aim for this processing to be reproducible. 
To aid in this regard scripts to reproduce the data reduction and and all figures based on these IRDIS data are included in the github repository for {\sc precision} as an example of how to use the code.

\subsection{Results}\label{sec:res}

\begin{figure*}
    \centering
    \includegraphics[width=\textwidth]{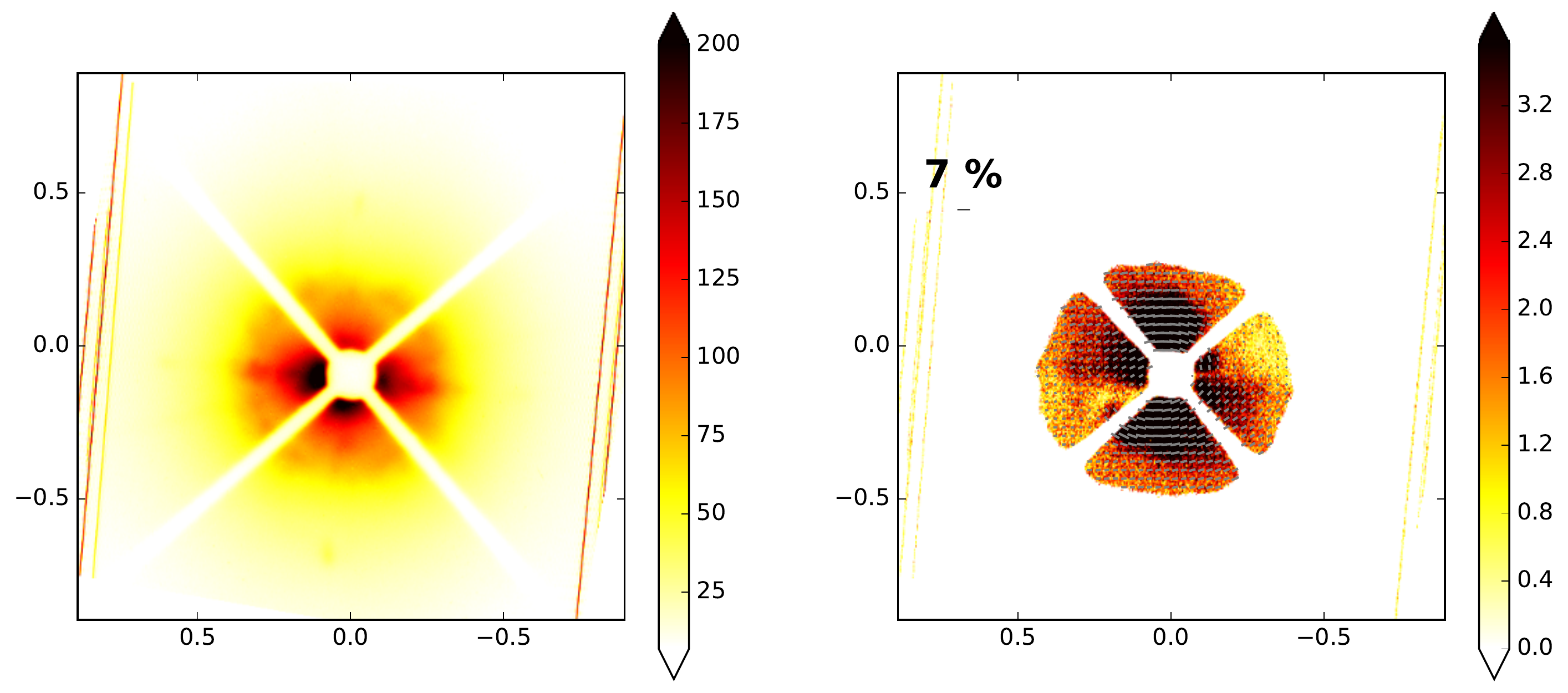}
    \includegraphics[width=\textwidth]{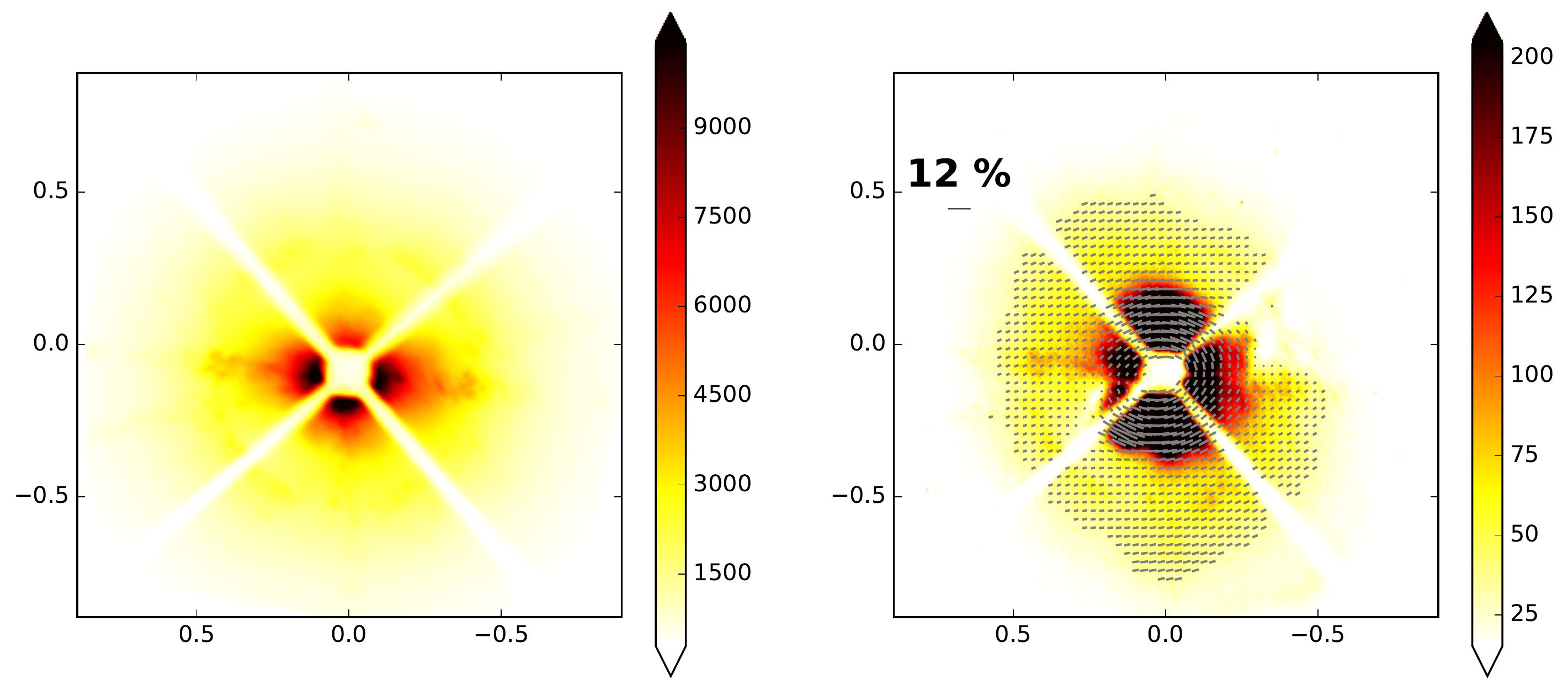}
    \caption{ZIMPOL imaging polarimetry of VX~Sgr. {\it Top}: V-band data, with total intensity on the left and polarised intensity on the right. Vectors are overlaid on the polarised-intensity plot to indicate the polarisation fraction and angle. The example vector in the upper-left corner shows a vector of the length of the maximum polarisation observed in the image. The stripes seen in the intensity image are a bias artefact which mostly cancels in the polarised intensity. Offsets are given in arcseconds, with North up and East to the left. {\it Bottom}: As above, for the Cnt820 filter.}
    \label{fig:zimvx}
\end{figure*}

Figure~\ref{fig:zimvx} shows the reduced intensity and polarised intensity images in both filters from the ZIMPOL observations. 
The total intensity shows little structure, with most of the emission probably arising from the stellar PSF and imperfect AO correction.
However, the polarisation images show a clear, extended signal, with the polarisation vectors producing the strong centro-symmetric pattern with polarisation fraction of several per cent, characteristic of scattering by dust in an axisymmetric envelope.
The significant ($\sim$ a factor of 2) elongation of the polarised intensity in the North-South direction argues against spherical symmetry, suggesting that we may be seeing an oblate spheroid, disc or torus at relatively low inclination.
Alternatively, the density could be enhanced in the North-South direction, with an otherwise symmetrical distribution. 
However, the higher polarisation fraction to the North and South of the source suggests that these regions are closer to the optimum angle for polarisation { \citep[typically $\sim$ 80 -- 120$^\circ$ and a weak function of wavelength, see e.g.][]{2007AJ....133.2730J}} than the rest of the envelope, consistent with an inclined non-spherical structure. 
While these data cannot conclusively identify the geometry, it is clear that there must be some degree of asymmetry, either in the form of a non-spherical or a clumpy outflow. 
No point-like sources are visible in the data.

\begin{figure*}
    \centering
    \includegraphics[width=1.05\textwidth,clip=true,trim=3cm 2.5cm 3cm 3cm]{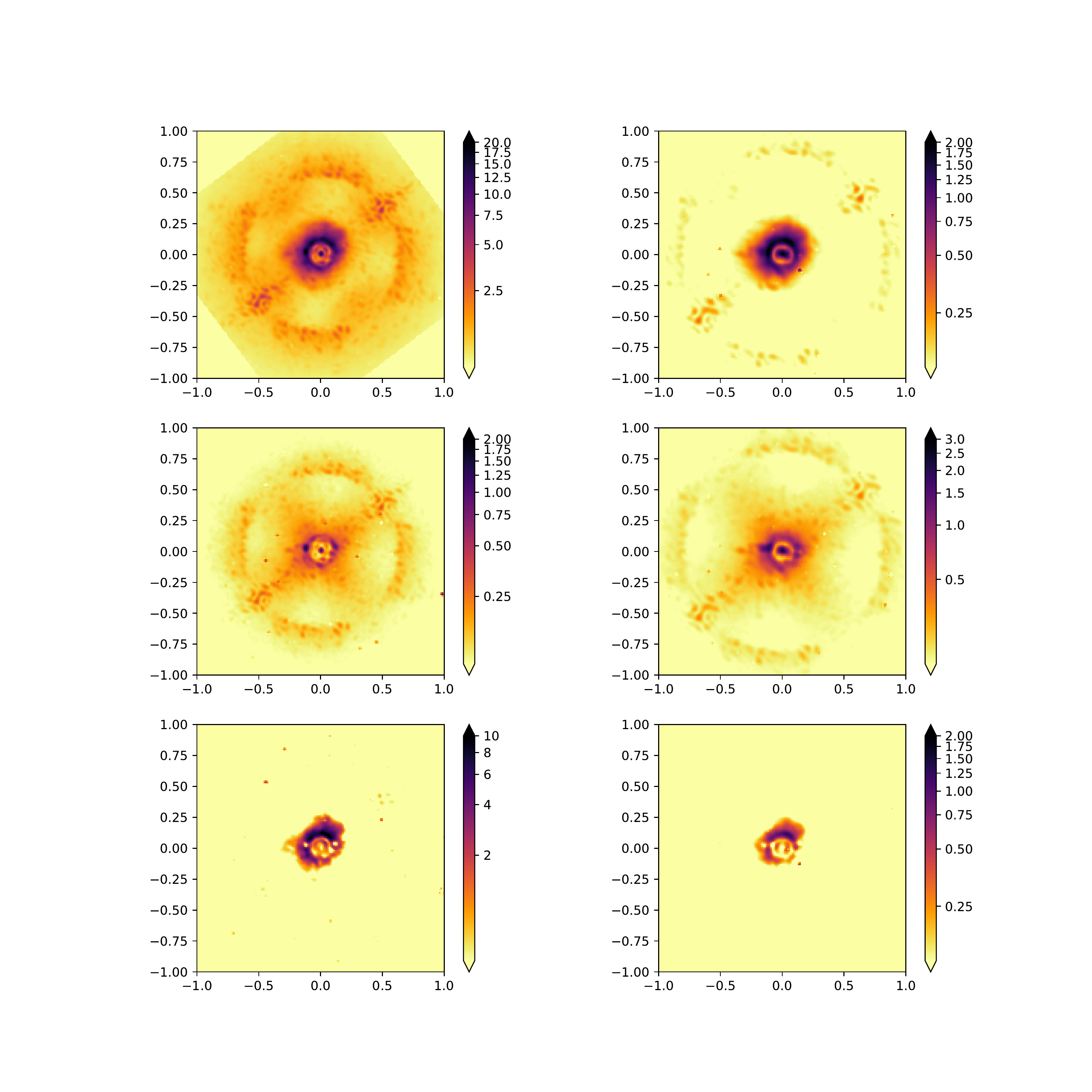} 
    \caption{IRDIS imaging of VX~Sgr and PSF reference, showing the central $2^{\prime\prime} \times 2^{\prime\prime}$ region of the field. Offsets are given in arc seconds, with North up and East to the left. The colour scale shows intensity with $\sinh^{-1}$ scaling to provide high dynamic range. The images on the left show the J-band data and those on the right the H-band images. {\it top}: Combined data of VX~Sgr only. Extended emission is clearly seen to the north and north-east of the star with the same shape and structure in both bands. The bright ring at $\sim 0\farcs5$ is an artefact caused by the finite control radius of the AO system ($\sim 20 \lambda/D$) - modes on angular scales larger than this remain uncorrected. {\it middle}: Combined data for the PSF reference 70~Sgr. No extended emission is visible. {\it bottom}: Images of VX~Sgr after scaled reference PSFs have been subtracted from to suppress artifacts related to the PSF and the finite AO control radius. The extended, asymmetric structure is still clearly visible in both bands, making it likely that this is real extended emission rather than flux leakage.}
    \label{fig:irdvx}
\end{figure*}

The IRDIS data are shown in Fig.~\ref{fig:irdvx}, which show an asymmetrical region of extended emission around the edge of the coronagraph in both filters.
This is visible both in the combined data and after the subtraction of the scaled reference PSF, although both the central spot and the bright ring at the AO control radius have been efficiently removed.
This region is substantially more extended than the PSF reference source, suggesting that this is NIR flux scattered from the circumstellar material rather than stellar flux leaking around the coronagraph due to poor centring or AO performance, and strongly implying that the asymmetry is real. 
The asymmetry of the emission is suggestive of the inner region of a disc or torus seen nearly face on, { consistent with the Northern sector of the outflow being closer to the observer and hence producing more flux through the difference in scattering phase function at different angles \citep[e.g.][]{Ysard2018},}  although a spherical shell cannot be excluded on the basis of this data alone.
Alternatively, the images may imply that there is significant clumpy or filamentary structure to the extended component, although additional, deeper observations would be required to determine whether these constitute discrete components in the ejecta or are artefacts.
In either case the data implies a significant deviation from spherical symmetry in the mass-loss geometry close to the star, and follow-up observations,  particularly non-coronagraphic imaging, is required to analyse these features further.
Similar to the optical data, no point-like sources are visible in the coadded classical images. 

{
Neither set of observations shows evidence that they are dominated by any of the known aberations that affect SPHERE data \citep{Cantalloube2019Msngr}.
Of these effects, only the ``wind-driven halo'' and ``low-wind effect'' are known to produce asymmetrical structures at times.
However, these distinctive effects can be excluded based on the morphology of the sources in our data.
}
\section{Discussion}
While neither the optical polarimetry nor the infrared imaging alone can conclusively identify the morphology of the emission, taken in combination they are suggestive of an equatorially-enhanced (disc- or torus-like)  circumstellar environment, rather than an expanding spherical shell.
{ The source is asymmetrical across the full wavelength range, suggestive of anisotropic (forward-dominated) scattering where one side of the object is slightly closer to the observer than the other. 
This is true not just in intensity, but polarised intensity as well, confirming that either the scattering angle or dust density must be changing with azimuthal angle.
Finally, while the mass-loss rate and optical polarisation are very high, the line of sight extinction is low and the star is visible in the optical, suggesting that we are viewing the system in a direction with relatively little circumstellar material, similar to the LMC red supergiant WOH\,G64 \citep{Ohnaka2008}.
Combined, these present a compelling, although not conclusive, argument in favour of a non-spherical shell.}

Similarly, \citet{2012ApJ...754...47S} found that SiO maser emission forms a ring with a very narrow range of line-of-sight velocities, while \citet{1994AJ....107.1469D} and \citet{2004ApJ...605..436M} used infrared interferometry to infer the presence of a centrosymmetric structure in dust emission with an inner radius of around 60 mas. 
\citet{2006AJ....131..603S} found a similar emission region to our data but did not see any significant asymmetry in the optical, attributing a slight deviation from circular symmetry to CCD bleeding. 
However, somewhat more extended faint emission is visible to the north in their figure 7, in the opposite direction to the CCD bleed they comment upon.
The asymmetry in H$_2$O masers seen by \citet{2018arXiv180704455Y} coincides with what is seen in our IRDIS images, confirming the asymmetry in the dust distribution.
If the NIR scattered light is assumed to come from the near-side of a moderately-inclined disc or torus, this is also consistent with the magnetic field orientation determined by \citet{2005A&A...434.1029V}.

However, as an apparently-single star, non-spherical geometry is difficult to explain, as there is no obvious source of angular momentum to produce a disc.
It is possible that any companion is too faint or too close to detect in our observations, or that VX~Sgr is already in a common-envelope phase, and hence could not be detected by \citet{2010A&A...511A..51C} in NIR interferometric observations. 
While one of the bright spots inferred by \citet{2010A&A...511A..51C} lies outside the photosphere, the fact that it is brighter at longer wavelengths implies that it is more likely to be structure in the ejecta than a companion.
On the other hand, if a companion had previously been accreted it might have spun the star up, providing the necessary rotational energy to eject or form a torus \citep[e.g.][]{1999ApJ...512..322C}.
Other possibilities include that the magnetic field inferred by \citet{2005A&A...434.1029V} is involved creating an underdense region near the poles, or that a bipolar outflow has evacuated the polar region.

The problem of non-spherical outflows is common to other hypergiants, most famously \VY. 
This may represent some similarity in other aspects, for example that their high mass-loss rates may be driven by intrinsically asymmetric processes \citet{2007AJ....133.2716H}.
In particular, these asymmetries can already be seen to have set in within $\sim$50\,$R_\ast$ in our data, suggesting that they are an initial condition rather than being imposed on the outflow at larger radii.
The ubiquity of inhomogeneities in all the RSGs over a wide parameter space that have been studied in sufficient detail to reveal them (e.g. $\alpha$~Ori, \citealt{2016A&A...585A..28K}, VY~CMa \citealt{2001AJ....121.1111S} and others) suggests that this may be a feature of RSG mass-loss, and bear information about the mechanisms intiating the outflow.


\begin{figure}
    \centering
    \includegraphics[width=0.5\textwidth,clip=true,trim=0cm 5cm 0cm 5cm]{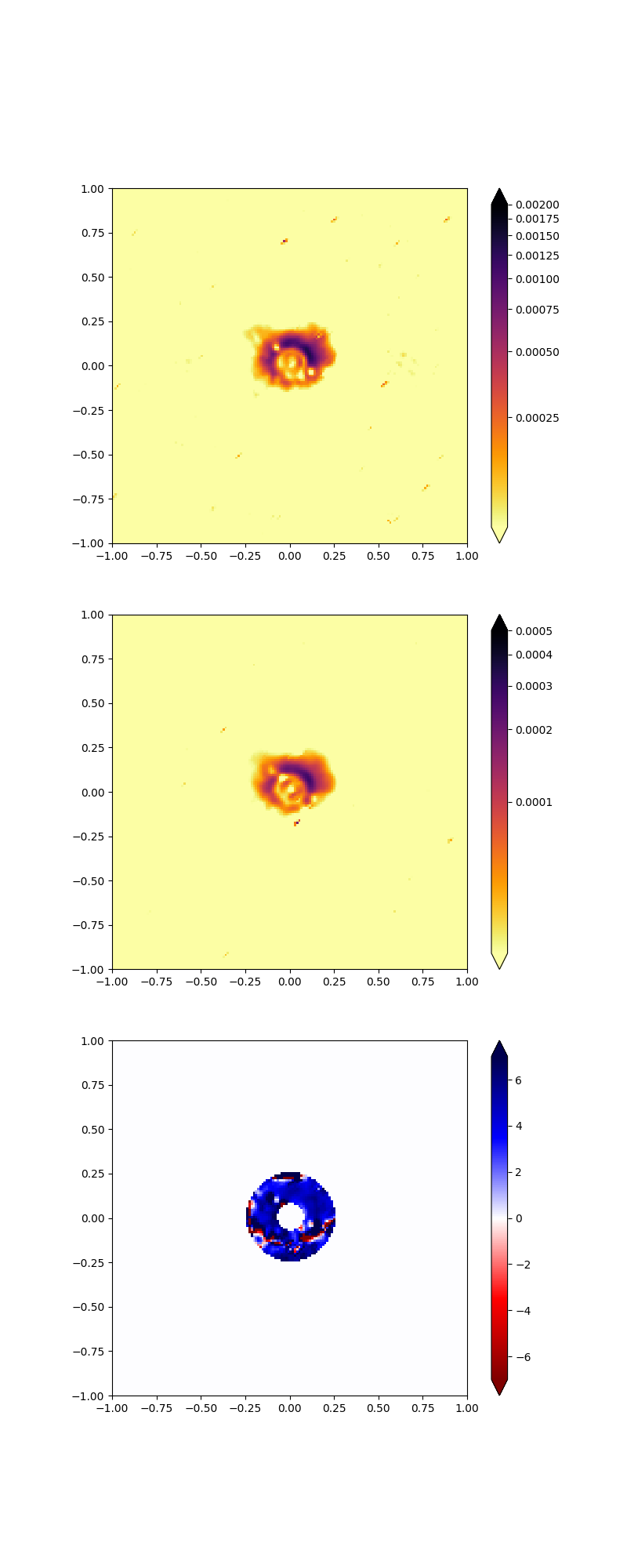}
    \caption{{\it Top}: Contrast-calibrated J-band image, shown in the detector frame. {\it Middle}: Contrast-calibrated H-band image. {\it Bottom}: Ratio map of the NIR PSF-subtracted, contrast-calibrated images, with blue corresponding to stronger J-band scattering. The area inside the coronagraph has been masked out, as has the area outside 250 mas, where the colour is dominated by the effect of the difference in resolution on the location of the residuals from PSF subtraction. }
    \label{fig:nircolourmap}
\end{figure}

In order to explore the properties of the dust producing the NIR scattered light, we produce a NIR colour map using the PSF-subtracted images (bottom row of Fig.~\ref{fig:irdvx}).
To remove the overall trend of the stellar colour, we first produce `contrast--calibrated' images by performing relative flux calibration (Sect.~\ref{sec:rfc}) to leave the `intrinsic' colour of the dust, which is directly related to the ratio of the scattering cross-sections at the observed wavelengths.
{ Contrary to expectations, the J$/$H flux ratio is $\sim 4$ in the region with significant extended emission, bluer than would be expected based on Rayleigh scattering, which should provide an upper limit to the flux ratio of $\left(\frac{1.24}{1.65}\right)^{-4}\sim 3$ if scattering is assumed to be isotropic.
This effect cannot be due to absorption by dust, as that would preferentially suppress the scattered light at J-band, lowering the flux ratio.
This may indicate that the scattering is significantly non-isotropic, as expected for typical astronomical dust.
}

It is also important to consider whether the variability of VX~Sgr could have any impact on our interpretation.
Our optical and near-IR observations are separated by approximately 3 months, which corresponds to slightly more than 10\% of the stars $\sim$700 day pulsation cycle \citep[e.g.][]{2010ApJS..190..203P}. 
While the typical optical-variability amplitude of VX~Sgr is several magnitudes, 10\% of the cycle is probably too short for this to make a significant impact, as the bolometric amplitude is much smaller.

Other sources of variability include dust formation and star spots. Depending on exactly when in the pulsation cycle our observations were obtained, it is possible that new dust may have formed between the two epochs, altering the illumination of the outer envelope. However, dust formation happens over the longer timescale of the pulsation period \citep[e.g.][]{2008A&A...491L...1H}, so the dust mass in the innermost outflow is unlikely to have changed significantly on this timescale. 
Starspots would change the illumination of the envelope as the star rotates, but without any knowledge of the rotation period of the star, it seems difficult to estimate the impact of this.

\section{Summary}

We present a new pipeline, \pipe, for reducing near-infrared high-contrast imaging observations.
At present, the pipeline supports reduction of data from the IRDIS module of VLT/SPHERE.
\pipe~is pure python, and designed to be fast, memory efficient and extensible.
To demonstrate this pipeline, it is applied to near-infrared high-contrast imaging observations of the extreme red supergiant VX~Sgr, which are compared with optical polarimetric imaging taken with ZIMPOL.

{ The SPHERE observations show clear evidence that the innermost outflow is non-spherical; the polarised intensity varies with azimuth by a factor of 2 while the near-infared scattered light is clumpy and asymmetrical.}
The ZIMPOL data for VX~Sgr show a centrosymmetric pattern characteristic of dust scattering from the surface of a centrally illuminated disc or torus, aligned with the rings of maser emission.
This is supported by the strong asymmetry seen in near infrared high-contrast imaging, which also shows some evidence of more complex structures.
These asymmetries correspond well with structures seen in water maser emission and with the inferred magnetic field direction.
This may indicate the recent onset of equatorially-enhanced mass loss, as has been inferred for \VY~and IRC+10420.
The similarities between these sources in other regards supports suggestions of a sequence, in which VX~Sgr is beginning a phase of enhanced mass loss similar to that recently experienced by \VY, and both will eventually evolve bluewards to become yellow hypergiants with complex circumstellar environments similar to IRC+10420.

The NIR colour and the ratio of optical polarisation fractions suggest that the dust in the outflow is dominated by large (several hundred nanometre) dust grains.
This adds to the growing weight of evidence that RSGs and O-rich evolved stars in general form large dust grains, and that radiation pressure from scattered light may play a role in driving the wind.

\section*{Acknowledgements}
We thank the anonymous referee for their careful reading of the manuscript and helpful suggestions which improved the paper. 
We wish to thank S. Moehler, J. Hron, K. Ohnaka, M. Takami, J. Karr and G. H.-M. Bertrang for helpful discussions and advice. 
We would also like to thank the ESO support astronomers for taking the data in service mode and our contact astronomers N. Gibson and M. van den Ancker for their help in preparing the observations and correcting issues. 
This research has been supported by grants MOST104-2628-M-001-004-MY3 and MOST107-2119-M-001-031-MY3 from the 
Ministry of Science and Technology
and
AS-IA-106-M03 from Academia Sinica.
This research has made use of the SIMBAD database, operated at CDS, Strasbourg, France \citep{2000Wenger}. 
This research has made use of NASA's Astrophysics Data System. 




\bibliographystyle{mnras}
\bibliography{biblio.bib} 








\bsp	
\label{lastpage}
\end{document}